\documentclass{article}%
\usepackage{amsfonts}
\usepackage{amsmath}
\usepackage{amssymb}
\usepackage{charter}
\usepackage{graphicx}%
\providecommand{\U}[1]{\protect \rule{.1in}{.1in}}

\RequirePackage{CJK}
\AtBeginDocument{\begin{CJK*}{GBK}{song}\CJKtilde}
\AtEndDocument{\end{CJK*}}
\begin{document}

\title{New 3-mode squeezing operator and squeezed vacuum state in 3-wave mixing
\thanks{{\small Work was supported by the National Natural Science Foundation
of China under grants 10775097 and the Key Programs Foundation of Ministry of
Education of \ China (No. 210115). }}}
\author{{\small Xue-xiang Xu}$^{1,2}$, {\small Hong-yi Fan}$^{1}$, {\small Li-yun
Hu}$^{2}${\small \thanks{E-mail:hlyun2008@126.com.},and Hong-chun Yuan}$^{1}$\\$^{1}${\small Department of Physics, Shanghai Jiao Tong University, Shanghai,
200240, China}\\$^{2}${\small College of Physics \& Communication Electronics, Jiangxi Normal
University, Nanchang 330022, China}}
\maketitle

\begin{abstract}
{\small In a 3-wave mixing process occurring in some nonlinear optical medium
when }$a_{1}^{\dagger}${\small mode interacts with both }$a_{2}^{\dagger}%
${\small mode and }$a_{3}^{\dagger}${\small mode, we theoretically study the
squeezing effect generated by the operator }$S_{3}\equiv \exp[\mu(a_{1}%
a_{2}-a_{1}^{\dagger}a_{2}^{\dagger})+\nu(a_{1}a_{3}-a_{1}^{\dagger}%
a_{3}^{\dagger})]${\small . The new 3-mode squeezed vacuum state in Fock space
is derived, and the uncertainty relation for it is demonstrated, It turns out
that }$S_{3}${\small may exhibit enhanced squeezing. By virtue of the
technique of integration within an ordered product (IWOP) of operators, we
also derive }$S_{3}${\small 's normally ordered expansion. The Wigner function
of new 3-mode squeezed vacuum state is calculated by using the Weyl ordering
invariance under similar transformations.}



\end{abstract}

\section{Introduction}

Nowadays quantum entanglement is the focus of quantum information research and
attracts many interests due to its wide applications in quantum communication
\cite{1,2}. Entangled states have brought much attention and interests of
physicists \cite{3,4}. The usual two-mode squeezed state, generated from a
parametric amplifier \cite{5}, not only exhibits squeezing, but also quantum
entanglement between the idle-mode and the signal-mode in frequency domain.
Therefore, it is simultaneously a typical entangled state of continuous
variable. Theoretically, the two-mode squeezed state is constructed by acting
a two-mode squeezing operator $S_{2}=\exp[\lambda(a_{1}a_{2}-a_{1}^{\dagger
}a_{2}^{\dagger})]$ \cite{6,7} on the two-mode vacuum state $\left \vert
00\right \rangle $, i.e. $S_{2}\left \vert 00\right \rangle =$sech$\lambda
\exp[-a_{1}^{\dagger}a_{2}^{\dagger}\tanh \lambda]\left \vert 00\right \rangle $,
where $\lambda \ $is a squeezing parameter, and $a_{i}$($a_{j}^{\dagger}$) Bose
annihilation (creation) operator satisfying $[a_{i},a_{j}^{\dagger}%
]=\delta_{ij}$. Using the relation between Bose operators ($a_{i}%
,a_{i}^{\dagger}$) and the coordinate and momentum operators%
\begin{equation}
Q_{i}=\frac{a_{i}+a_{i}^{\dagger}}{\sqrt{2}},\ P_{i}=\frac{a_{i}%
-a_{i}^{\dagger}}{\sqrt{2}\mathtt{i}}, \label{0.1}%
\end{equation}
one can recast $S_{2}$ into the form%
\begin{equation}
S_{2}=\exp \left[  \mathtt{i}\lambda \left(  Q_{1}P_{2}+Q_{2}P_{1}\right)
\right]  , \label{0.2}%
\end{equation}
noting%
\begin{equation}%
\begin{array}
[c]{c}%
\left[  Q_{1}P_{2},Q_{2}P_{1}\right]  =\mathtt{i}\left(  Q_{2}P_{2}-Q_{1}%
P_{1}\right)  ,\\
\left[  Q_{1}P_{2},\mathtt{i}\left(  Q_{2}P_{2}-Q_{1}P_{1}\right)  \right]
=2Q_{1}P_{2},\\
\left[  Q_{2}P_{1},\mathtt{i}\left(  Q_{2}P_{2}-Q_{1}P_{1}\right)  \right]
=-2Q_{2}P_{1},
\end{array}
\label{0.3}%
\end{equation}
thus there involves a $SU(1,1)$ algebraic stricture. In the state
$S_{2}\left \vert 00\right \rangle $, the variances of the two-mode quadrature
operators of light field,%
\begin{equation}
\mathfrak{X}=\frac{Q_{1}+Q_{2}}{2},\text{ }\mathfrak{P}=\frac{P_{1}+P_{2}}{2},
\label{0.4}%
\end{equation}
satisfying the commutation relation $[\mathfrak{X},\mathfrak{P}]=\frac
{\mathtt{i}}{2}$, exhibiting the standard squeezing, i.e.,%
\begin{equation}
\left \langle 00\right \vert S_{2}^{\dagger}\mathfrak{X}^{2}S_{2}\left \vert
00\right \rangle =\frac{1}{4}e^{-2\lambda},\left \langle 00\right \vert
S_{2}^{\dagger}\mathfrak{P}^{2}S_{2}\left \vert 00\right \rangle =\frac{1}%
{4}e^{2\lambda}, \label{0.5}%
\end{equation}
which satisfy $(\Delta \mathfrak{X})(\Delta \mathfrak{P})=\frac{1}{4}$.

An interesting question naturally arises: if $a_{1}^{\dagger}$ mode in a
nonlinear optical medium, interacting with both $a_{2}^{\dagger}$ mode and
$a_{3}^{\dagger}$ mode (e.g., a three-wave mixing), and the corresponding
three-mode exponential operator is introduced as%
\begin{equation}
S_{3}\equiv \exp[\mu(a_{1}a_{2}-a_{1}^{\dagger}a_{2}^{\dagger})+\nu(a_{1}%
a_{3}-a_{1}^{\dagger}a_{3}^{\dagger})]. \label{0.6}%
\end{equation}
Using Eq.(\ref{0.1}) we can recast $S_{3}$ into the form
\begin{equation}
S_{3}\equiv \exp \left[  \mathtt{i}\mu \left(  Q_{2}P_{1}+Q_{1}P_{2}\right)
+\mathtt{i}\nu \left(  Q_{3}P_{1}+Q_{1}P_{3}\right)  \right]  , \label{0.7}%
\end{equation}
where $\mu,$ $\nu$ are two different interaction parameters, then what is its
squeezing effect for the 3-mode quadratures of light field?

To answer this question we must know what is the state $S_{3}\left \vert
000\right \rangle $ ($\left \vert 000\right \rangle $ is the 3-mode vacuum state)
in Fock space, for this aim, we should know what is the normally ordered
expansion of $S_{3}$. But how to disentangle the exponential operator $S_{3}?$
Because there is no simple $SU(1,1)$ algebraic structure among $Q_{2}%
P_{1},Q_{1}P_{2},Q_{3}P_{1}$ and $Q_{1}P_{3},$ the disentangling seems hard.
Thus we turn to appeal to Dirac's coordinate representation and the technique
of integration within an ordered product (IWOP) of operators \cite{8,9,10,11}
to solve this problem. Our work is arranged as follows: firstly we derive the
explicit form of$\ S_{3}\left \vert 000\right \rangle ,$ then we demonstrate
that it really satisfies the Heisenberg uncertainty relation and may exhibit
squeezing enhancement. We also employ the technique of integration within an
ordered product (IWOP) of operators to derive the normally ordered expansion
of $S_{3}$. The Wigner function of $S_{3}\left \vert 000\right \rangle $ is
calculated by using the Weyl ordering invariance under similar transformations
\cite{12,13,14}.

\section{New 3-mode squeezed vacuum state}

For the sake of convenience, we rewrite $S_{3}$\ in Eq.(\ref{0.7}) as the
following compact form,%
\begin{equation}
S_{3}=\exp[\mathtt{i}Q_{i}\Lambda_{ij}P_{j}],i,j=1,2,3, \label{1.1}%
\end{equation}
where the repeated indices imply the Einstein summation notation, and%
\begin{equation}
\Lambda=\left(
\begin{array}
[c]{ccc}%
0 & \mu & \nu \\
\mu & 0 & 0\\
\nu & 0 & 0
\end{array}
\right)  , \label{1.2}%
\end{equation}
thus%
\begin{equation}
e^{\Lambda}=\allowbreak \left(
\begin{array}
[c]{ccc}%
\cosh r & \cos \theta \sinh r & \sin \theta \sinh r\\
\cos \theta \sinh r & \sin^{2}\theta+\cos^{2}\theta \cosh r & \frac{\sin2\theta
}{2}\left(  \cosh r-1\right) \\
\sin \theta \sinh r & \frac{\sin2\theta}{2}\left(  \cosh r-1\right)  & \sin
^{2}\theta \cosh r+\cos^{2}\theta
\end{array}
\right)  , \label{1.3}%
\end{equation}
its inverse is%
\begin{equation}
e^{-\Lambda}\allowbreak=\allowbreak \left(
\begin{array}
[c]{ccc}%
\cosh r & -\cos \theta \sinh r & -\sin \theta \sinh r\\
-\cos \theta \sinh r & \sin^{2}\theta+\cos^{2}\theta \cosh r & \frac{\sin
2\theta \left(  \cosh r-1\right)  }{2}\\
-\sin \theta \sinh r & \frac{\sin2\theta \left(  \cosh r-1\right)  }{2} &
\sin^{2}\theta \cosh r+\cos^{2}\theta
\end{array}
\right)  , \label{1.4}%
\end{equation}
where we have set%
\begin{equation}
r=\sqrt{\mu^{2}+\nu^{2}},\cos \theta=\frac{\mu}{r},\sin \theta=\frac{\nu}{r},
\label{1.5}%
\end{equation}
noting that $\Lambda$ is a symmetric matrix. Using the Baker-Hausdorff
formula,%
\begin{align}
e^{A}Be^{-A}  &  =B+\left[  A,B\right]  +\frac{1}{2!}\left[  A,\left[
A,B\right]  \right] \nonumber \\
&  +\frac{1}{3!}\left[  A,\left[  A,\left[  A,B\right]  \right]  \right]
+\cdots, \label{1.6}%
\end{align}
we see that $S_{3}$\ causes the following transformation%
\begin{equation}
S_{3}^{-1}Q_{k}S_{3}=(e^{-\Lambda})_{ki}Q_{i},\text{\ }S_{3}^{-1}P_{k}%
S_{3}=(e^{\Lambda})_{ki}P_{i}. \label{1.7}%
\end{equation}
It then follows that $S_{3}^{-1}a_{k}S_{3}=(e^{-\lambda \Lambda})_{ki}a_{i}$,
i.e.%
\begin{align}
S_{3}^{-1}a_{1}S_{3}  &  =a_{1}\cosh r-a_{2}^{\dag}\cos \theta \sinh
r-a_{3}^{\dag}\sin \theta \sinh r,\nonumber \\
S_{3}^{-1}a_{2}S_{3}  &  =-a_{1}^{\dag}\cos \theta \sinh r+a_{2}\left(  \sin
^{2}\theta+\cos^{2}\theta \cosh r\right) \nonumber \\
&  +\frac{1}{2}a_{3}\left(  \cosh r-1\right)  \sin2\theta,\label{1.8}\\
S_{3}^{-1}a_{3}S_{3}  &  =-a_{1}^{\dag}\sin \theta \sinh r+\frac{1}{2}%
a_{2}\left(  \cosh r-1\right)  \sin2\theta \nonumber \\
&  +a_{3}\left(  \sin^{2}\theta \cosh r+\cos^{2}\theta \right)  .\nonumber
\end{align}
Noticing that $S_{3}^{\dagger}=S_{3}^{-1}$ and $S_{3}^{\dagger}\left(  \mu
,\nu \right)  =S_{3}\left(  -\mu,-\nu \right)  $, from Eq.(\ref{1.8}) we also
have%
\begin{align}
S_{3}a_{1}S_{3}^{-1}  &  =a_{1}\cosh r+a_{2}^{\dag}\cos \theta \sinh
r+a_{3}^{\dag}\sin \theta \sinh r,\nonumber \\
S_{3}a_{2}S_{3}^{-1}  &  =a_{1}^{\dag}\cos \theta \sinh r+a_{2}\left(  \sin
^{2}\theta+\cos^{2}\theta \cosh r\right) \nonumber \\
&  +\frac{1}{2}a_{3}\left(  \cosh r-1\right)  \sin2\theta \label{1.9}\\
S_{3}a_{3}S_{3}^{-1}  &  =a_{1}^{\dag}\sin \theta \sinh r+\frac{1}{2}%
a_{2}\left(  \cosh r-1\right)  \sin2\theta \nonumber \\
&  +a_{3}\left(  \sin^{2}\theta \cosh r+\cos^{2}\theta \right)  .\nonumber
\end{align}

For convenience to write, we set $S_{3}\left \vert 000\right \rangle =\left \Vert
000\right \rangle $. In order to obtain the explicit form of $\left \Vert
000\right \rangle $, using Eq.(\ref{1.8}) and $a_{1}\left \vert 000\right \rangle
=0$, we operate $a_{1}$ on $\left \Vert 000\right \rangle $ and obtain%
\begin{align}
a_{1}\left \Vert 000\right \rangle  &  =S_{3}S_{3}^{-1}a_{1}S_{3}\left \vert
000\right \rangle \nonumber \\
&  =S_{3}(a_{1}\cosh r-a_{2}^{\dag}\cos \theta \sinh r-a_{3}^{\dag}\sin
\theta \sinh r)\left \vert 000\right \rangle \nonumber \\
&  =S_{3}(-a_{2}^{\dag}\cos \theta \sinh r-a_{3}^{\dag}\sin \theta \sinh
r)S_{3}^{-1}S_{3}\left \vert 000\right \rangle \nonumber \\
&  =-S_{3}(a_{2}^{\dag}\cos \theta \sinh r+a_{3}^{\dag}\sin \theta \sinh
r)S_{3}^{-1}\left \Vert 000\right \rangle , \label{1.10}%
\end{align}
then we continue to use Eq.(\ref{1.9}) to derive%
\begin{align}
a_{1}\left \Vert 000\right \rangle  &  =-\{[a_{1}\cos \theta \sinh r+a_{2}%
^{\dagger}\left(  \sin^{2}\theta+\cos^{2}\theta \cosh r\right) \nonumber \\
&  +\frac{1}{2}a_{3}^{\dagger}\left(  \cosh r-1\right)  \sin2\theta]\cos
\theta \sinh r\nonumber \\
&  +[a_{1}\sin \theta \sinh r+\frac{1}{2}a_{2}^{\dagger}\left(  \cosh
r-1\right)  \sin2\theta \nonumber \\
&  +a_{3}^{\dagger}\left(  \sin^{2}\theta \cosh r+\cos^{2}\theta \right)
]\sin \theta \sinh r\} \left \Vert 000\right \rangle \nonumber \\
&  =-(a_{1}\sinh^{2}r+a_{2}^{\dagger}\allowbreak \cos \theta \cosh r\sinh
r\nonumber \\
&  +\frac{1}{2}a_{3}^{\dagger}\sin \theta \sinh2r)\left \Vert 000\right \rangle ,
\label{1.11}%
\end{align}
so we reach the equation%
\begin{equation}
a_{1}\left \Vert 000\right \rangle =-\tanh r(a_{2}^{\dagger}\allowbreak
\cos \theta+a_{3}^{\dagger}\sin \theta)\left \Vert 000\right \rangle .
\label{1.12}%
\end{equation}
Similarly, operating $a_{2}$ on $\left \Vert 000\right \rangle $ and using
Eqs.(\ref{1.8}) and (\ref{1.9}) yields
\begin{align}
a_{2}\left \Vert 000\right \rangle  &  =S_{3}S_{3}^{-1}a_{2}S_{3}\left \vert
000\right \rangle =S_{3}(-a_{1}^{\dag}\cos \theta \sinh r)\left \vert
000\right \rangle \nonumber \\
&  =S_{3}(-a_{1}^{\dag}\cos \theta \sinh r)S_{3}^{-1}\left \Vert 000\right \rangle
\nonumber \\
&  =-(a_{1}^{\dag}\cosh r+a_{2}\cos \theta \sinh r\nonumber \\
&  +a_{3}\sin \theta \sinh r)\cos \theta \sinh r\left \Vert 000\right \rangle ,
\label{1.13}%
\end{align}
which leads to%
\begin{align}
&  [a_{2}\left(  1+\cos^{2}\theta \sinh^{2}r\right)  +\frac{1}{2}a_{3}%
\sin2\theta \sinh^{2}r]\left \Vert 000\right \rangle \nonumber \\
&  =-\frac{1}{2}a_{1}^{\dag}\cos \theta \sinh2r\left \Vert 000\right \rangle .
\label{1.14}%
\end{align}
On the other hand, operating $a_{3}$ on $\left \Vert 000\right \rangle $ and
using Eqs.(\ref{1.8}) and (\ref{1.9}) yields
\begin{align}
a_{3}\left \Vert 000\right \rangle  &  =S_{3}S_{3}^{-1}a_{3}S_{3}\left \vert
000\right \rangle =S_{3}(-a_{1}^{\dag}\sin \theta \sinh r)\left \vert
000\right \rangle \nonumber \\
&  =S_{3}(-a_{1}^{\dag}\sin \theta \sinh r)S_{3}^{-1}\left \Vert 000\right \rangle
\nonumber \\
&  =-(a_{1}^{\dag}\cosh r+a_{2}\cos \theta \sinh r\nonumber \\
&  +a_{3}\sin \theta \sinh r)\sin \theta \sinh r\left \Vert 000\right \rangle ,
\label{1.15}%
\end{align}
i.e.,
\begin{align}
&  [a_{3}\left(  1+\sin^{2}\theta \sinh^{2}r\right)  +\frac{1}{2}a_{2}%
\sin2\theta \sinh^{2}r]\left \Vert 000\right \rangle \nonumber \\
&  =-\frac{1}{2}a_{1}^{\dag}\sin \theta \sinh2r\left \Vert 000\right \rangle .
\label{1.16}%
\end{align}
Combining Eqs.(\ref{1.14}) and (\ref{1.16}) we have%
\begin{equation}
a_{2}\left \Vert 000\right \rangle =-a_{1}^{\dag}\tanh r\cos \theta \left \Vert
000\right \rangle , \label{1.17}%
\end{equation}
and%
\begin{equation}
a_{3}\left \Vert 000\right \rangle =-a_{1}^{\dag}\tanh r\sin \theta \left \Vert
000\right \rangle . \label{1.18}%
\end{equation}
From Eqs.(\ref{1.12}),(\ref{1.17}) and (\ref{1.18}), we may predict that
$\left \Vert 000\right \rangle $\ has the following explicit form:%
\begin{equation}
\left \Vert 000\right \rangle =N\exp[-(a_{2}^{\dag}\cos \theta+a_{3}^{\dag}%
\sin \theta)a_{1}^{\dag}\tanh r]\left \vert 000\right \rangle , \label{1.19}%
\end{equation}
where $N$ is the normalization constant, which can be \ determined by
$\left \langle 000\right.  \left \Vert 000\right \rangle =1$, and we calculate
$N=\sec$h$r.$

\section{Squeezing property and quantum fluctuation in $\left \Vert
000\right \rangle $}

Squeezing is an important phenomenon in quantum theory and has many
applications in various areas in quantum optics and quantum information
\cite{15}. In this section, we examine the quadrature squeezing effects of
$\left \Vert 000\right \rangle $. The quadratures in the 3-mode case are defined
as%
\begin{equation}
X_{1}=\frac{1}{\sqrt{6}}\sum_{i=1}^{3}Q_{i},\text{ }X_{2}=\frac{1}{\sqrt{6}%
}\sum_{i=1}^{3}P_{i}, \label{2.1}%
\end{equation}
which satisfy the relation $[X_{1},X_{2}]=\frac{\mathtt{i}}{2}.$ Their
variances are $\left(  \Delta X_{i}\right)  ^{2}=\left \langle X_{i}%
^{2}\right \rangle -\left \langle X_{i}\right \rangle ^{2}$, $i=1,2.$ Noting the
expectation values of $X_{1}$ and $X_{2}$ in the state $\left \Vert
000\right \rangle $ is $\left \langle X_{1}\right \rangle =\left \langle
X_{2}\right \rangle =0$. With the help of Eq.(\ref{1.8}), we can calculate that
the corresponding variances in the state $\left \Vert 000\right \rangle $:
(noting $\Lambda$ is symmetric)
\begin{align}
\left(  \triangle X_{1}\right)  ^{2}  &  =\left \langle 000\right \vert
S_{3}^{-1}X_{1}^{2}S_{3}\left \vert 000\right \rangle \nonumber \\
&  =\frac{1}{6}\sum_{i=1}^{3}\sum_{j=1}^{3}(e^{-\Lambda})_{ki}(e^{-\Lambda
})_{jl}\left \langle 000\right \vert Q_{k}Q_{l}\left \vert 000\right \rangle
\nonumber \\
&  =\frac{1}{12}\sum_{i=1}^{3}\sum_{j=1}^{3}(e^{-\Lambda})_{ki}(e^{-\Lambda
})_{jl}\left \langle 000\right \vert a_{k}a_{l}^{\dagger}\left \vert
000\right \rangle \nonumber \\
&  =\frac{1}{12}\sum_{i=1}^{3}\sum_{j=1}^{3}(e^{-\Lambda})_{ki}(e^{-\Lambda
})_{jl}\delta_{kl}\nonumber \\
&  =\frac{1}{12}\underset{i,j}{\sum^{3}}(e^{-2\Lambda})_{ij}, \label{2.2}%
\end{align}
and%
\begin{equation}
\left(  \triangle X_{2}\right)  ^{2}=\left \langle 000\right \vert S_{3}%
^{-1}X_{2}^{2}S_{3}\left \vert 000\right \rangle =\frac{1}{12}\underset
{i,j}{\sum^{3}}(e^{2\Lambda})_{ij}. \label{2.3}%
\end{equation}
The explicit form of the matrices $e^{2\Lambda}$ and $e^{-2\Lambda}$ can be
derived from Eq.(\ref{1.3}) and (\ref{1.4}), so we can obtain%
\begin{align}
\left(  \triangle X_{1}\right)  ^{2}  &  =\frac{1}{12}[(2\cosh2r+1)+\sin
2\theta(\allowbreak \cosh2r-1)\nonumber \\
&  +2\left(  \cos \theta+\sin \theta \right)  \sinh2r], \label{2.4}%
\end{align}
and%
\begin{align}
\left(  \triangle X_{2}\right)  ^{2}  &  =\frac{1}{12}[(2\cosh2r+1)+\sin
2\theta(\allowbreak \cosh2r-1)\nonumber \\
&  -2\left(  \cos \theta+\sin \theta \right)  \sinh2r]. \label{2.5}%
\end{align}
We can successfully verify
\begin{align}
&  (\triangle X_{1})(\triangle X_{2})\nonumber \\
&  =\frac{1}{12}\sqrt{\left(  4\cosh2r+4\right)  +\left(  1-2\sinh^{2}%
r\sin2\theta \right)  ^{2}}\nonumber \\
&  \geqslant \frac{1}{12}\sqrt{\left(  4\cosh2r+4\right)  +\left(  1-2\sinh
^{2}r\right)  ^{2}}\nonumber \\
&  =\frac{1}{12}\sqrt{\frac{1}{2}\cosh4r+\frac{17}{2}}\geqslant \frac{1}{4},
\label{2.6}%
\end{align}
which confirms the uncertainty relation of quantum mechanics.

To see the trend of squeezing effects in the $X_{1}-$ or $X_{2}-$direction, we
plot $\left(  \Delta X_{1}\right)  ^{2}$ and $\left(  \Delta X_{2}\right)
^{2}$\ as the function of parameter $\mu$ for different $\nu$ in Fig.1. When
$\nu=0,$ it exhibits the usual two mode squeezing effect depending on the
varying $\mu$, $\left(  \Delta X_{1}\right)  ^{2}$ increases accompanying
$\left(  \Delta X_{2}\right)  ^{2}$ decreases; when $\nu=0.5$, $\left(  \Delta
X_{1}\right)  ^{2}\ $increases more than the case of $\nu=0$, which exhibits
enhanced squeezing in certain domain of $\mu$. In Fig.2, we plot the
uncertainty value $(\triangle X_{1})(\triangle X_{2})$\ as the function of $r$
for different $\theta$.

\section{Normally ordered form of $S_{3}$}

We calculate the normally ordered form (denoted by $\colon \colon$) of $S_{3}$
by inserting the completeness relation of coherent state
\begin{equation}
S_{3}=\int \frac{d^{2}z_{1}d^{2}z_{2}d^{2}z_{3}}{\pi^{3}}S_{3}\left \vert
z_{1}z_{2}z_{3}\right \rangle \left \langle z_{1}z_{2}z_{3}\right \vert .
\label{3.1}%
\end{equation}
where $\left \vert z_{1}z_{2}z_{3}\right \rangle $\ is the three-mode coherent
state and $\left \vert z_{i}\right \rangle =\exp[-\frac{\left \vert
z_{i}\right \vert ^{2}}{2}+z_{i}a_{i}^{\dag}]\left \vert 0_{i}\right \rangle $,
$i=1,2,3$.

Using the relations in Eqs.(\ref{1.8}) and (\ref{1.9}), we have $\allowbreak
$the explicit relation of $S_{3}\left \vert z_{1}z_{2}z_{3}\right \rangle
\allowbreak$%
\begin{align}
&  S_{3}\left \vert z_{1}z_{2}z_{3}\right \rangle \nonumber \\
&  =\exp(-\sum_{i=1}^{3}\frac{\left \vert z_{i}\right \vert ^{2}}{2})S_{3}%
\exp \left(  z_{1}a_{1}^{\dagger}+z_{2}a_{2}^{\dagger}+z_{3}a_{3}^{\dagger
}\right)  S_{3}^{-1}S_{3}\left \vert 000\right \rangle \nonumber \\
&  =\frac{1}{\cosh r}\exp(-\sum_{i=1}\frac{\left \vert z_{i}\right \vert ^{2}%
}{2}+z_{1}z_{2}\cos \theta \tanh r+z_{1}z_{3}\sin \theta \tanh r)\nonumber \\
&  \times \exp \allowbreak \{ \frac{1}{\cosh r}[a_{1}^{\dag}z_{1}+\allowbreak
(a_{2}^{\dag}\left(  \sin^{2}\theta \cosh r+\cos^{2}\theta \right)  -\frac{1}%
{2}a_{3}^{\dag}\left(  \cosh r-1\right)  \sin2\theta)z_{2}\nonumber \\
&  +(a_{3}^{\dag}\left(  \sin^{2}\theta+\cos^{2}\theta \cosh r\right)
-\frac{1}{2}a_{2}^{\dag}\left(  \cosh r-1\right)  \sin2\theta)z_{3}%
]\} \nonumber \\
&  \times \exp[-a_{1}^{\dag}\tanh r(a_{2}^{\dag}\cos \theta+a_{3}^{\dag}%
\sin \theta)]\left \vert 000\right \rangle . \label{3.2}%
\end{align}
Substituting Eq.(\ref{3.2}) into Eq.(\ref{3.1}), noticing that $\left \vert
000\right \rangle \left \langle 000\right \vert =\colon \exp(-a_{1}^{\dag}%
a_{1}-a_{2}^{\dag}a_{2}-a_{3}^{\dag}a_{3})\colon$, and using the following
formula%
\begin{equation}
\int \frac{d^{2}z}{\pi}\exp \left(  \zeta \left \vert z\right \vert ^{2}+\xi z+\eta
z^{\ast}\right)  =-\frac{1}{\zeta}e^{-\frac{\xi \eta}{\zeta}},\text{
\  \ }\mathtt{Re}\left(  \zeta \right)  <0, \label{3.4}%
\end{equation}
as well as the IWOP technique, we can obtain the explicit normally ordered
expansion of $S_{3}$:
\begin{align}
S_{3}  &  =\frac{1}{\cosh r}\exp[-a_{1}^{\dag}(a_{2}^{\dag}\cos \theta
+a_{3}^{\dag}\sin \theta)\tanh r]\nonumber \\
&  \times \colon \exp[\frac{1-\cosh r}{\cosh r}(a_{1}^{\dag}a_{1}+a_{2}^{\dag
}a_{2}\cos^{2}\theta \nonumber \\
&  +a_{3}^{\dag}a_{3}\sin^{2}\theta+\frac{1}{2}a_{2}a_{3}^{\dag}\sin
2\theta+\frac{1}{2}a_{2}^{\dag}a_{3}\sin2\theta)]\colon \nonumber \\
&  \times \exp[a_{1}(a_{2}\cos \theta+\allowbreak a_{1}\sin \theta)\tanh r].
\label{3.5}%
\end{align}

\section{Wigner function of $\left \Vert 000\right \rangle $}

Wigner distribution function of quantum states \cite{16,17,18} is widely
studied in quantum statistics and quantum optics and is very important tool
for a global description of nonclassical effect in the quantum system, which
can be measured by various means such as photon counting experiment and
homodyne tomography. Now we derive the Wigner function of $\left \Vert
000\right \rangle $ by using a new method.

Recalling that in Ref. \cite{14} we have introduced the Weyl ordering form of
single-mode Wigner operator $\Delta_{1}\left(  q_{1},p_{1}\right)  $,%
\begin{equation}
\Delta_{1}\left(  q_{1},p_{1}\right)  =%
\genfrac{}{}{0pt}{}{:}{:}%
\delta \left(  q_{1}-Q_{1}\right)  \delta \left(  p_{1}-P_{1}\right)
\genfrac{}{}{0pt}{}{:}{:}%
, \label{4.1}%
\end{equation}
where the symbols$%
\genfrac{}{}{0pt}{}{:}{:}%
\genfrac{}{}{0pt}{}{:}{:}%
$ denote the Weyl ordering, while its normal ordering form is%
\begin{equation}
\Delta_{1}\left(  q_{1},p_{1}\right)  =\frac{1}{\pi}\colon \exp \left[  -\left(
q_{1}-Q_{1}\right)  ^{2}-\left(  p_{1}-P_{1}\right)  ^{2}\right]  \colon.
\label{4.2}%
\end{equation}
Thus the Wigner function for $\left \vert 0\right \rangle $ can be easily
expressed as $\left \langle 0\right \vert \Delta_{1}\left(  q_{1},p_{1}\right)
\left \vert 0\right \rangle =\frac{1}{\pi}\exp(-q_{1}^{2}-p_{1}^{2})$. Note that
the order of Bose operators $a_{1}$ and $a_{1}^{\dagger}$ within a normally
ordered product (or a Weyl ordered product) can be permuted. That is to say,
even though $[a_{1},a_{1}^{\dagger}]=1$, we can have $\colon a_{1}%
a_{1}^{\dagger}\colon=\colon a_{1}^{\dagger}a_{1}\colon$ and$%
\genfrac{}{}{0pt}{}{:}{:}%
a_{1}a_{1}^{\dagger}%
\genfrac{}{}{0pt}{}{:}{:}%
=%
\genfrac{}{}{0pt}{}{:}{:}%
a_{1}^{\dagger}a_{1}%
\genfrac{}{}{0pt}{}{:}{:}%
.$ The Weyl ordering of operators has a remarkable property, i.e., the
Weyl-ordering invariance of operators under similar transformations, which
means%
\begin{equation}
U%
\genfrac{}{}{0pt}{}{:}{:}%
\left(  \circ \circ \circ \right)
\genfrac{}{}{0pt}{}{:}{:}%
U^{-1}=%
\genfrac{}{}{0pt}{}{:}{:}%
U\left(  \circ \circ \circ \right)  U^{-1}%
\genfrac{}{}{0pt}{}{:}{:}%
, \label{4.3}%
\end{equation}
as if the \textquotedblleft fence" $%
\genfrac{}{}{0pt}{}{:}{:}%
\genfrac{}{}{0pt}{}{:}{:}%
$did not exist when $U$ operates.

For 3-mode case, the Weyl ordering form of the Wigner operator is
\begin{equation}
\Delta_{3}\left(  \mathbf{q},\mathbf{p}\right)  =%
\genfrac{}{}{0pt}{}{:}{:}%
\delta \left(  \mathbf{q}-\mathbf{Q}\right)  \delta \left(  \mathbf{p}%
-\mathbf{P}\right)
\genfrac{}{}{0pt}{}{:}{:}%
, \label{4.4}%
\end{equation}
where $\mathbf{Q}=(Q_{1},Q_{2},Q_{3})^{T}$, $\mathbf{P}=(P_{1},P_{2}%
,P_{3})^{T}$, $\mathbf{q}=(q_{1},q_{2},q_{3})^{T}$ and $\mathbf{p}%
=(p_{1},p_{2},p_{3})^{T}$. Then according to the Weyl ordering invariance
under similar transformations and using Eq.(\ref{1.7}), we have%
\begin{align}
&  S_{3}^{-1}\Delta_{3}\left(  \mathbf{q},\mathbf{p}\right)  S_{3}\nonumber \\
&  =S_{3}^{-1}%
\genfrac{}{}{0pt}{}{:}{:}%
\delta \left(  \mathbf{q}-\mathbf{Q}\right)  \delta \left(  \mathbf{p}%
-\mathbf{P}\right)
\genfrac{}{}{0pt}{}{:}{:}%
S_{3}\nonumber \\
&  =%
\genfrac{}{}{0pt}{}{:}{:}%
\delta \left(  q_{k}-(e^{-\Lambda})_{ki}Q_{i}\right)  \delta \left(
p_{k}-(e^{\Lambda})_{ki}P_{i}\right)
\genfrac{}{}{0pt}{}{:}{:}%
\nonumber \\
&  =%
\genfrac{}{}{0pt}{}{:}{:}%
\delta \left(  (e^{\Lambda})_{ki}q_{i}-Q_{k}\right)  \delta \left(
(e^{-\Lambda})_{ki}p_{i}-P_{k}\right)
\genfrac{}{}{0pt}{}{:}{:}%
\nonumber \\
&  =%
\genfrac{}{}{0pt}{}{:}{:}%
\delta \left(  \mathbf{q}^{\prime}-\mathbf{Q}\right)  \delta \left(
\mathbf{p}^{\prime}-\mathbf{P}\right)
\genfrac{}{}{0pt}{}{:}{:}%
\nonumber \\
&  =\Delta_{3}\left(  \mathbf{q}^{\prime},\mathbf{p}^{\prime}\right)  ,
\label{4.5}%
\end{align}
where $q_{k}^{\prime}=(e^{\Lambda})_{ki}q_{i},$ $p_{k}^{\prime}=(e^{-\Lambda
})_{ki}p_{i}$.$\allowbreak$ Thus the Wigner function of $\left \Vert
000\right \rangle $ is%
\begin{align}
&  \left \langle 000\right \vert S_{3}^{-1}\Delta_{3}\left(  \mathbf{q}%
,\mathbf{p}\right)  S_{3}\left \vert 000\right \rangle \nonumber \\
&  =\left \langle 000\right \vert \Delta_{3}\left(  \mathbf{q}^{\prime
},\mathbf{p}^{\prime}\right)  \left \vert 000\right \rangle \nonumber \\
&  =\frac{1}{\pi^{3}}\exp \left(  -\mathbf{q}^{T}e^{2\Lambda}\mathbf{q}%
-\mathbf{p}^{T}e^{-2\Lambda}\mathbf{p}\right)  , \label{4.6}%
\end{align}
where $e^{2\Lambda}$ and $e^{-2\Lambda}$ are given by $e^{\Lambda}$ in
Eq.(\ref{1.3}) and $e^{-\Lambda}$ in Eq.(\ref{1.4}), respectively.

In summary, we have shown that the operator $S_{3}\equiv \exp[\mu(a_{1}%
a_{2}-a_{1}^{\dagger}a_{2}^{\dagger})+\nu(a_{1}a_{3}-a_{1}^{\dagger}%
a_{3}^{\dagger})]$ is a new 3-mode squeezed operator by calculating the
quantum fluctuation for 3-mode quadratures. We have obtained the new 3-mode
squeezed vacuum state and derived the normally ordered expansion of $S_{3}$.
The IWOP technique brings convenience in our derivation.

\bigskip

\newpage

\bigskip \begin{figure}[ptb]
\label{Fig1}
\centering \includegraphics[width=8cm]{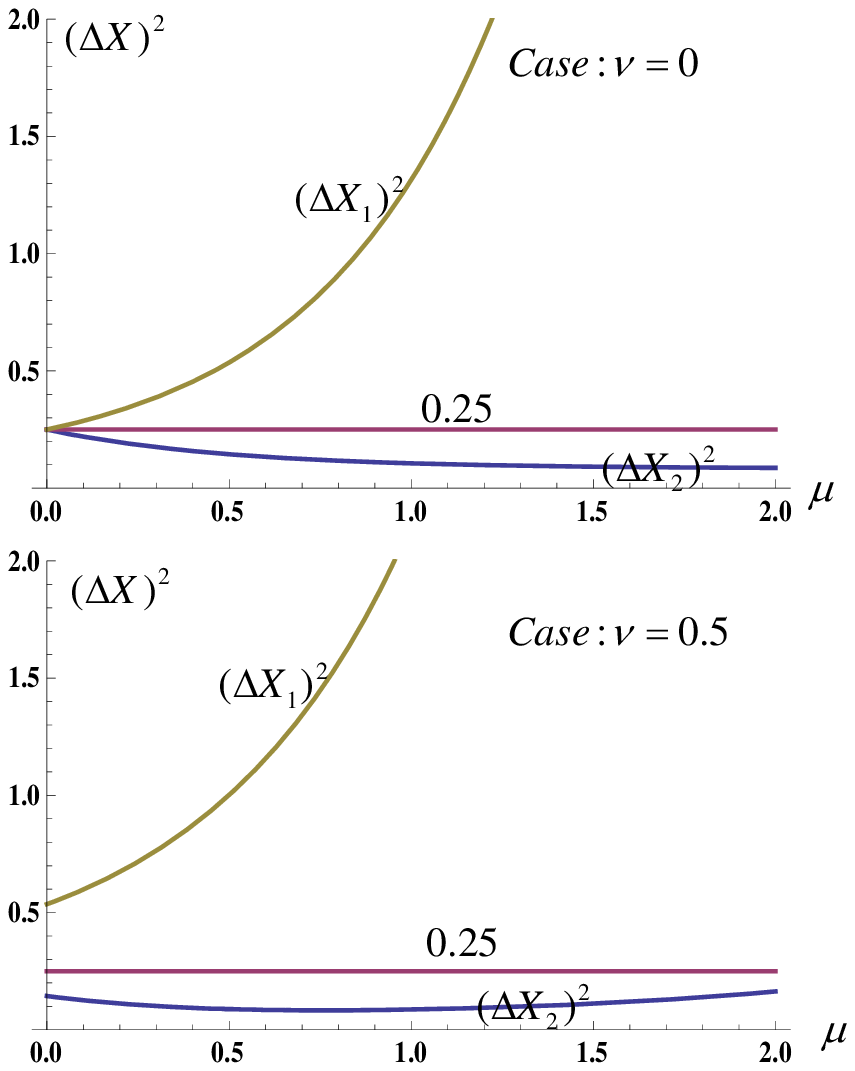}\caption{(Colour online) The
quantity $(\triangle X_{1})^{2}$ and $(\triangle X_{2})^{2}$ as the function
of squeezing parameter $\mu$\ for different case $\nu=0$ and $\nu=0.5$.}%
\end{figure}\begin{figure}[ptb]
\label{Fig2}
\centering \includegraphics[width=8cm]{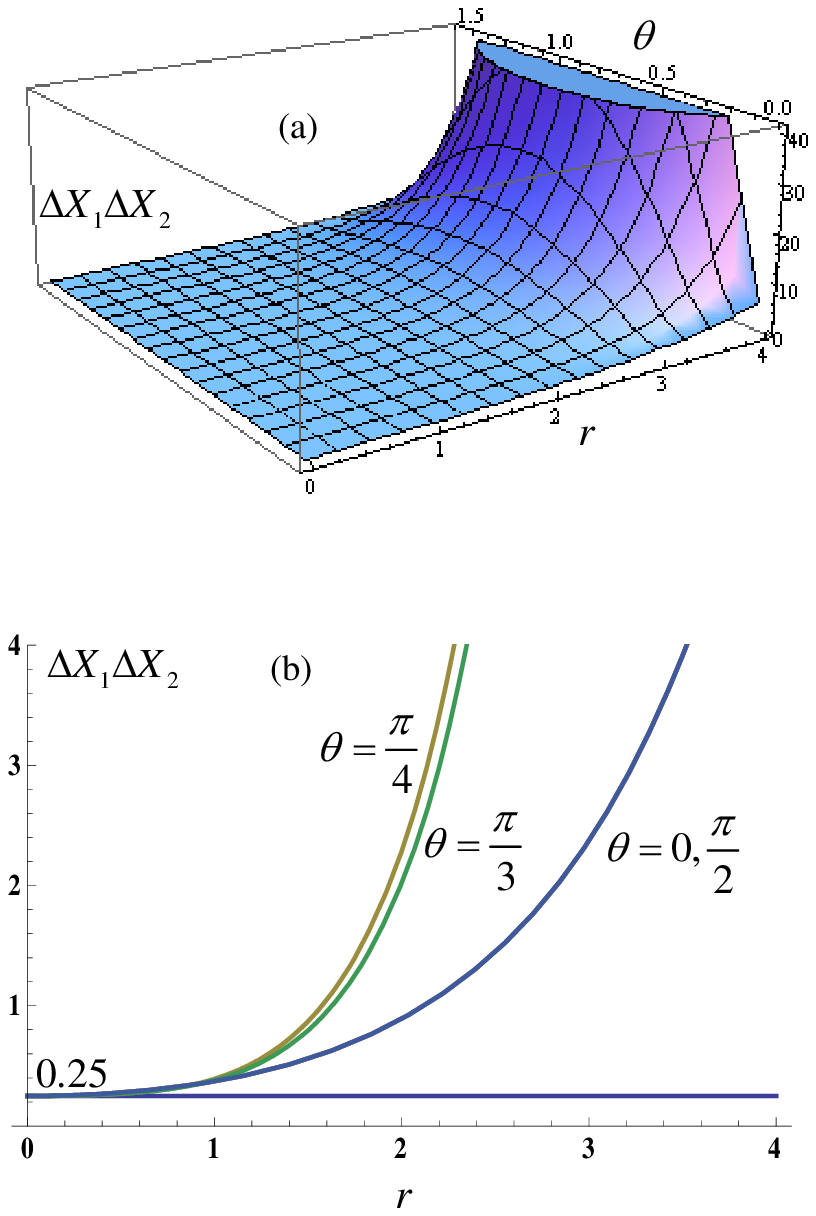}\caption{(Colour online) The
uncertainty value $(\triangle X_{1})(\triangle X_{2})$\ as the function of $r$
for different $\theta$.}%
\end{figure}

\end{document}